\documentclass[aps,prl,twocolumn,superscriptaddress,showpacs]{revtex4-1}

\usepackage{graphicx}
\begin{document}

\title{Dependence of Dzyaloshinskii-Moriya interaction on the oxygen coverage in Pt/Co/MOx trilayers}

\author{Dayane de Souza Chaves}
\affiliation{Univ.~Grenoble Alpes, CNRS, Institut N\'eel, F-38000 Grenoble, France}
\author{Fernando Ajejas}
\affiliation{Universidad Aut\'{o}noma de Madrid Campus de Cantoblanco, 28049, Madrid, Spain
}
\author{Viola Krizakova}
\affiliation{Univ.~Grenoble Alpes, CNRS, Institut N\'eel, F-38000 Grenoble, France}
\author{Jan Vogel}
\affiliation{Univ.~Grenoble Alpes, CNRS, Institut N\'eel, F-38000 Grenoble, France}
\author{Stefania Pizzini}
\affiliation{Univ.~Grenoble Alpes, CNRS, Institut N\'eel, F-38000 Grenoble, France}\email{stefania.pizzini@neel.cnrs.fr}

\begin{abstract}
We have studied the interfacial Dzyaloshinskii-Moriya interaction (DMI) in a series of Pt/Co/MOx (M=Al, Gd) trilayers in which the degree of oxidation of the top Co interface is varied. To access to reliable values of the DMI strength, we have used a method based on the measurement of the saturation velocity of field driven chiral N\'{e}el domain walls. We show that the effective DMI strength in the Pt/Co/MOx trilayers varies with the oxidation degree of the Co/MOx interface. This strongly suggests that the Co/MOx interface gives a distinct contribution to the total DMI, adding to that of the Pt/Co interface. The DMI presents a maximum for the oxygen coverage maximizing also the interface magnetic anisotropy energy $K_{s}$. This calls for common microscopic origins for the contributions of the Co/MOx interface to DMI and $K_{s}$.
\end{abstract}

\maketitle

In the last few years, the study of magnetic multilayers has known a strong revival, following the discovery that the interfacial Dzyaloshinskii-Moriya interaction (DMI)\cite{Dzyaloshinskii1957,Moriya1960} may become non-negligible with respect to the Heisenberg exchange and lead to non-trivial magnetic configurations. In ultrathin ferromagnetic (FM) layers in contact with a heavy metal (HM) with large spin-orbit coupling (SOC) within a non-centrosymmetric stack, the interfacial DMI may lead to the stabilization of chiral magnetic textures such as chiral N\'{e}el walls \cite{Thiaville2012} and magnetic skyrmions \cite{Skyrme1960, Fert2017}. The chirality of the spin configuration introduces new features of the magnetization dynamics. Large DMI stabilizes the domain wall (DW) internal structure against precession \cite{Thiaville2012}, so that DWs can be driven to very large velocities by magnetic fields \cite{Yoshimura2015,Yamada2015,Pham2016}. Moreover, chiral N\'{e}el walls and skyrmions can be displaced efficiently  when driven by spin polarized currents \textit{via} spin-orbit torques \cite{Emori2013,Ryu2013,Fert2017}. Routes to optimize the DMI are therefore targeted, as both kinds of magnetic textures are forecast to become carriers of binary information in future spintronic devices, like racetrack memories \cite{Parkin2008,Fert2013}.

The understanding of the microscopic mechanisms leading to interfacial DMI - and therefore to chiral magnetic textures - is still elusive, and little is known about the relation between the DMI and the details of the electronic structure. The DMI was first introduced as a super-exchange term in low-symmetry crystals \cite{Dzyaloshinskii1957,Moriya1960}, later extended to metals \cite{Levy1980}. In 1990 Fert showed with an analytical model that the DMI term is allowed also in thin films due to the symmetry reduction at the interfaces \cite{Fert1990}. At FM/HM interfaces the super-exchange between FM spins is mediated by the heavy metal atoms and the DMI strength is expected to scale with the HM spin-orbit coupling. Recently,
a few groups have developed Density Functional Theory (DFT) calculations to address quantitatively the DMI in several HM/FM bilayer systems \cite{Freimuth2014,Yang2015,Belabbes2016} and models to predict the variation of the DMI for $3d$/$5d$ interfaces have been proposed \cite{Kashid2014,Belabbes2016}. Such calculations can guide experimentalists towards the optimization of materials with large DMI, although they consider ideal interfaces, while experiments concentrate on multilayer systems grown by magnetron sputtering, where poly-crystallinity and interfacial roughness influence the strength and even the sign of the DMI \cite{Wells2017}. Experimental techniques to measure DMI are based on the non-reciprocal dispersion of spin waves within the FM layers \cite{Di2015,Belmeguenai2015,Cho2015,Tacchi2017}, on the anisotropic propagation of chiral N\'{e}el walls \cite{Je2013,Hrabec2014,Lavrijsen2015,Vanatka2015} or on chiral nucleation of reversed domains \cite{Pizzini2014}.

Among the many systems with DMI, the Pt/Co interface is the most studied, following the original work by Miron \textit{et al.} that found very large current-driven domain wall velocities in Pt/Co/AlOx trilayers \cite{Miron2011}. \textit{Ab initio} \cite{Yang2015,Yang2016,Freimuth2014} and experimental studies on polycrystalline stacks \cite{Pizzini2014,Belmeguenai2015,Cho2015,Kim2016} agree on the fact that such an interface is the source of strong DMI favoring homochiral magnetic textures (chiral N\'{e}el walls and skyrmions) with anticlockwise rotation of the magnetic moments .

The contribution of the FM/oxide interface to the total (effective) DMI in HM/FM/MOx trilayers is also the subject of large interest, as the absence of strong SOC calls for a new microscopic mechanism. Recent \textit{ab initio} calculations  have predicted that the sign and the strength of the DMI in Ir/Fe/O trilayers can be tuned by varying the oxygen coverage of the Fe layer \cite{Belabbes2017}. Similar results are anticipated in the case of any electronegative atom covering the cobalt layer and this suggests that the DMI could be electrically tuned. Strong contributions to the total DMI coming from Co/O \cite{Boulle2016,Freimuth2014,Yang2016} and Co/C \cite{Yang2016,Yang-Chen2017} interfaces in Pt/Co/MgO or Pt/Co/graphene trilayers were also predicted. The modification of the $3d$ band structure at the Fermi level \cite{Belabbes2017}, or a Rashba type DMI are invoked \cite{Yang2016,Yang-Chen2017}. However such effects have not been clearly confirmed up to now, as it is not straightforward experimentally to separate the DMI contributions coming from the bottom and top FM interfaces.

In this paper, we report the study of interfacial DMI in a series of Pt/Co/MOx  samples in which the degree of oxidation of the top Co interface is varied. To access to reliable values of the DMI strength, we have introduced an original method based on the measurement of the saturation velocity of field driven chiral N\'{e}el domain walls \cite{Pham2016}. We show that the effective DMI strength in the Pt/Co/MOx trilayers varies with the oxidation degree of the cobalt layer, and presents a maximum for the condition where the magnetic anisotropy energy is also maximum.

\begin{figure}
  \begin{center}
    \includegraphics[width=6cm]{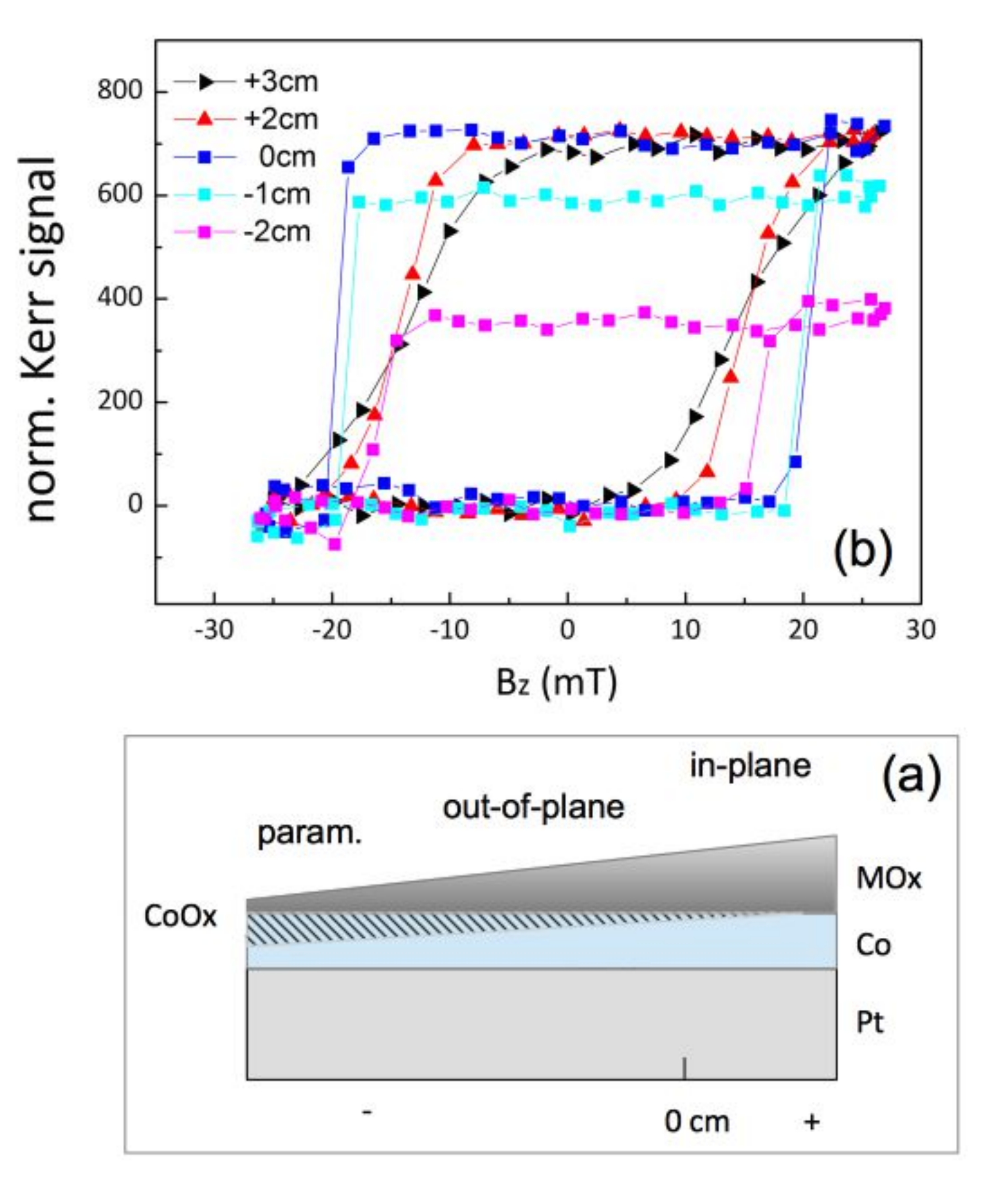}
  \end{center}
\caption{ (a)  Sketch of the sample stacks: the AlOx (GdOx) layer thickness increases going from negative to positive x. The samples are 8cm long and we have arbitrarily designed x=0 as the position around which the PMA is maximum; (b) hysteresis loops measured by magneto-optical Kerr effect in different positions of the Pt/Co/AlOx wedge.}
\label{fig:Figure-1}
\end{figure}

\begin{figure}[t]
  \begin{center}
    \includegraphics[width=8.5cm]{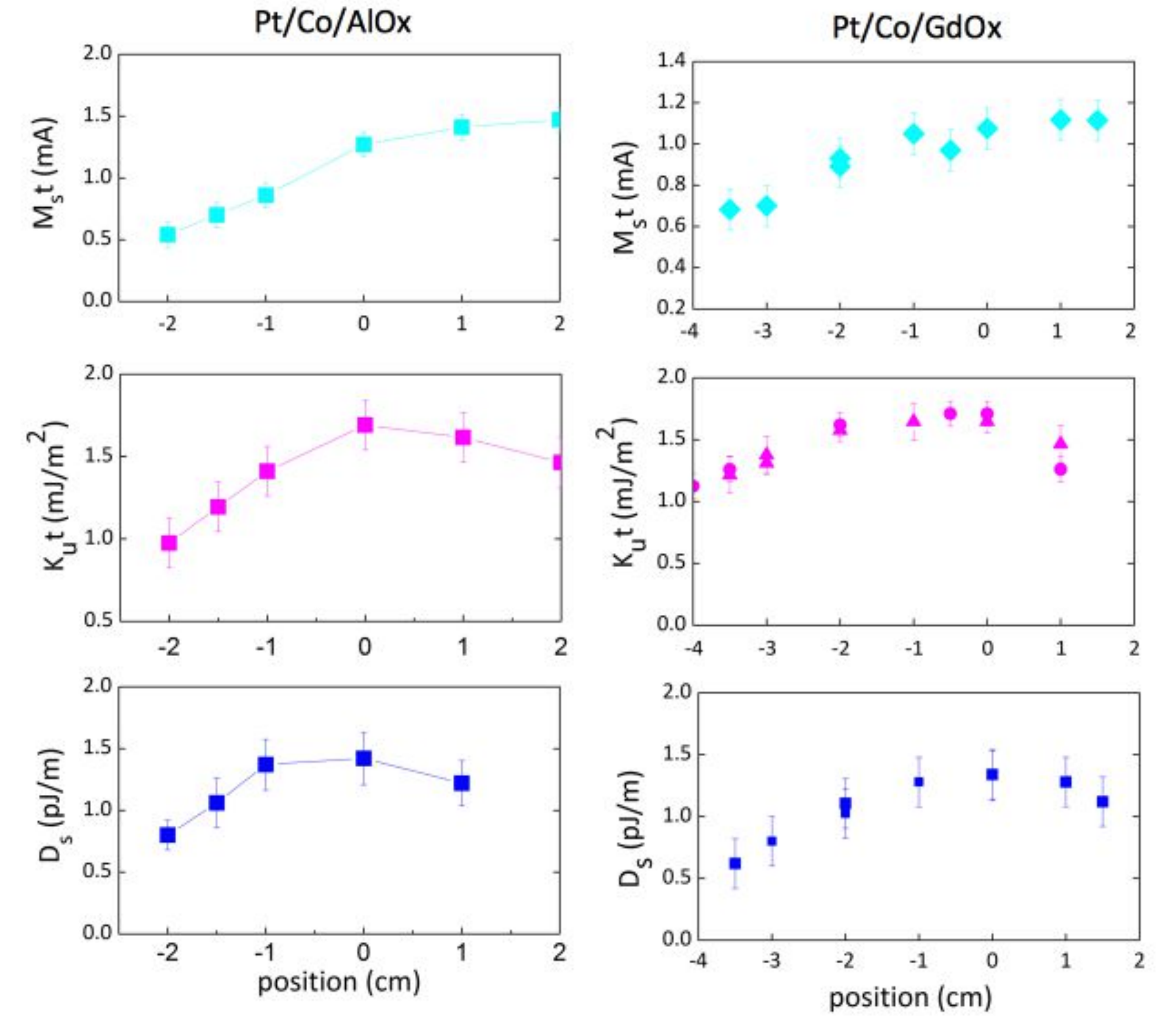}
  \end{center}
\caption{Unit surface magnetization $M_{s}t$, interface magnetic anisotropy energy $K_{u}t$ and interfacial DMI strength $D_{s}$ measured as a function of wedge position (as a function of oxidation) for the Pt/Co/AlOX sample (left) and the Pt/Co/GdOx sample (right) . }
\label{fig:Figure-4}
\end{figure}

Measurements were carried out on Pt(30)/ Co(0.8)/AlOx(1-3)/Pt(2) and Pt(4)/Co(1)/GdOx(1-3)/Al(3) stacks (thickness in nm). All samples were grown on Si/SiO$_{2}$ substrates by magnetron sputtering at room temperature. By choosing an off-axis geometry between target and substrate, the  Al and Gd layers were grown in the shape of a wedge with a thickness gradient. The Al and Gd  layers were then oxidised with an oxygen plasma. The thickness gradient results in a variable oxidation of these layers, and as a consequence in a different rate of oxygen covering the top Co interface (see sketch in Fig.~\ref{fig:Figure-1}(a)).
The magnetic properties of the two samples were obtained from VSM-SQUID measurements and magneto-optical Kerr microscopy in different positions of the wedge corresponding to different oxidation conditions. The unit surface magnetization $M_{s}t$ ($t$ being the thickness of the un-oxidized Co layer) and the in-plane saturation field $\mu_{0}H_{K}$ are reported in Table 1. The $M_{s}t$ dependence on sample position is also plotted for the two samples in Fig.~\ref{fig:Figure-4}.  The hysteresis loops measured by polar Kerr microscopy for the Pt/Co/AlOx are shown in Fig.~\ref{fig:Figure-1}(b). Within the region studied in this work, the magnetization of both Pt/Co/AlOx and Pt/Co/GdOx  is out-of-plane. The unit surface magnetization decreases going towards thinner AlOx (GdOx) layers, as the Co layer becomes more and more oxidized. This is also shown by the hysteresis loops, where the decrease  of the Kerr signal going towards thinner Al layers is related to the decrease of the magnetic Co thickness. X-ray reflectivity data indicate that with the thinnest AlOx (or GdOx) layers the 1~nm-thick Co layer is oxidized over $\sim$0.3~nm. VSM-SQUID measurements carried out at low temperature show that the thin oxide layer does not induce exchange bias and that the Curie temperature is well above room temperature even for the most oxidized part of the Co layer.

\begin{figure} [b]
  \begin{center}
    \includegraphics[width=8cm]{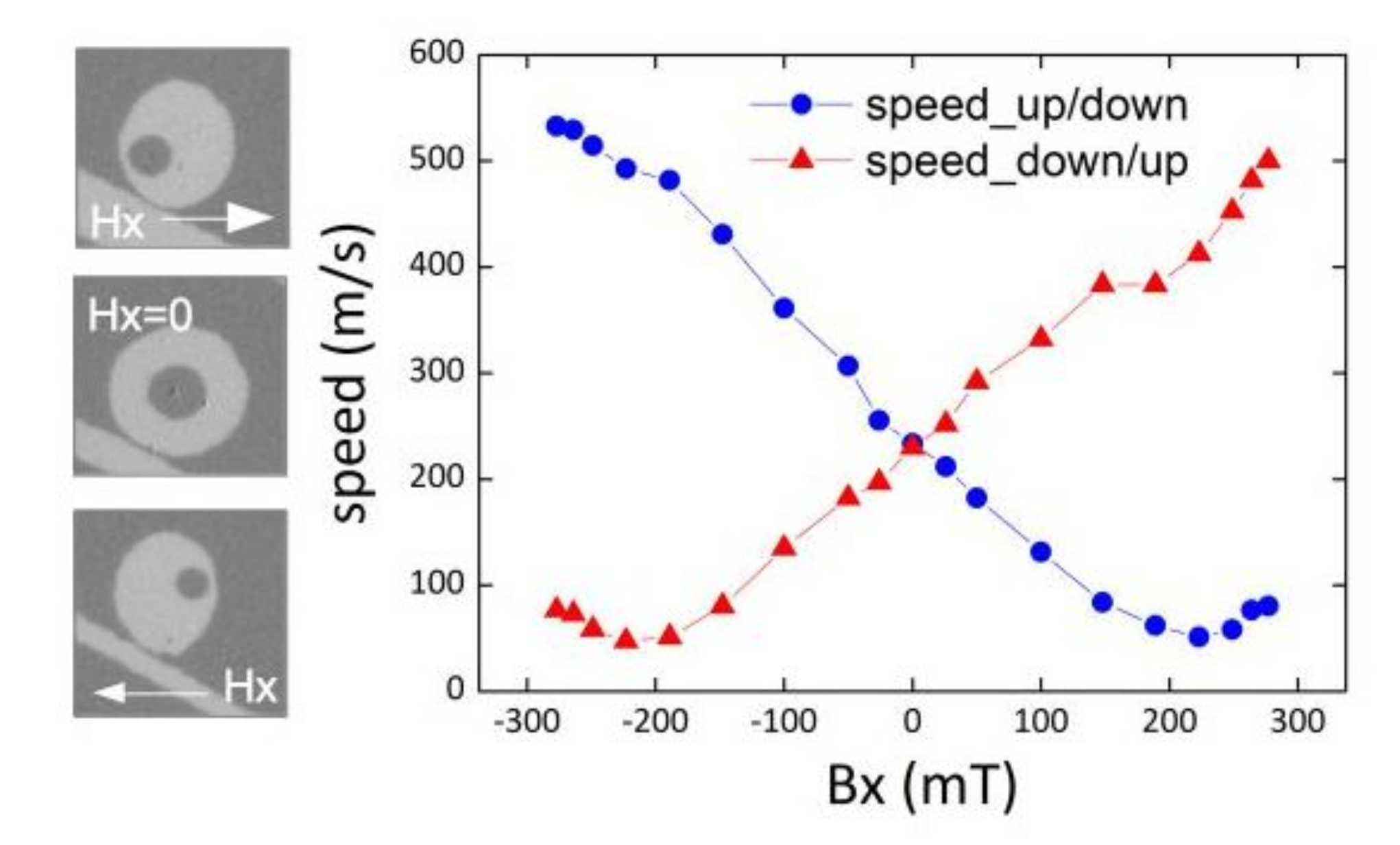}
  \end{center}
\caption{DW speed \textit{vs.} in-plane magnetic field $B_{x}$, measured for up/down and down/up DW propagating in $\pm$x direction, driven by 30ns long $B_{z}$ pulses of 80 mT , for the Pt/Co/GdOx sample in position +0.5cm. The differential Kerr images represent the expansion of a bubble domain driven by $B_{z}$ pulses with constant $B_{x}$ = -120~mT, 0~mT and +120~mT}
\label{fig:Figure-3}
\end{figure}

The domain wall velocities were measured in the continuous films, using wide field magneto-optical Kerr microscopy. The film magnetization was first saturated in the out-of-plane direction. An opposite  magnetic field pulse $H_{z}$ was then applied to nucleate one or several reverse domains. The DW velocity was deduced from the expansion of the initial bubble domain, after the application of further magnetic field pulses (see an example in Fig.~\ref{fig:Figure-3}). Using a 200~$\mu$m diameter coil associated to a fast current source, out-of-plane magnetic field pulses  of strength up to $\mu_0H_{z}$=700 mT and duration down to 20~ns were applied.

\begin{figure}
  \begin{center}
    \includegraphics[width=8.5cm]{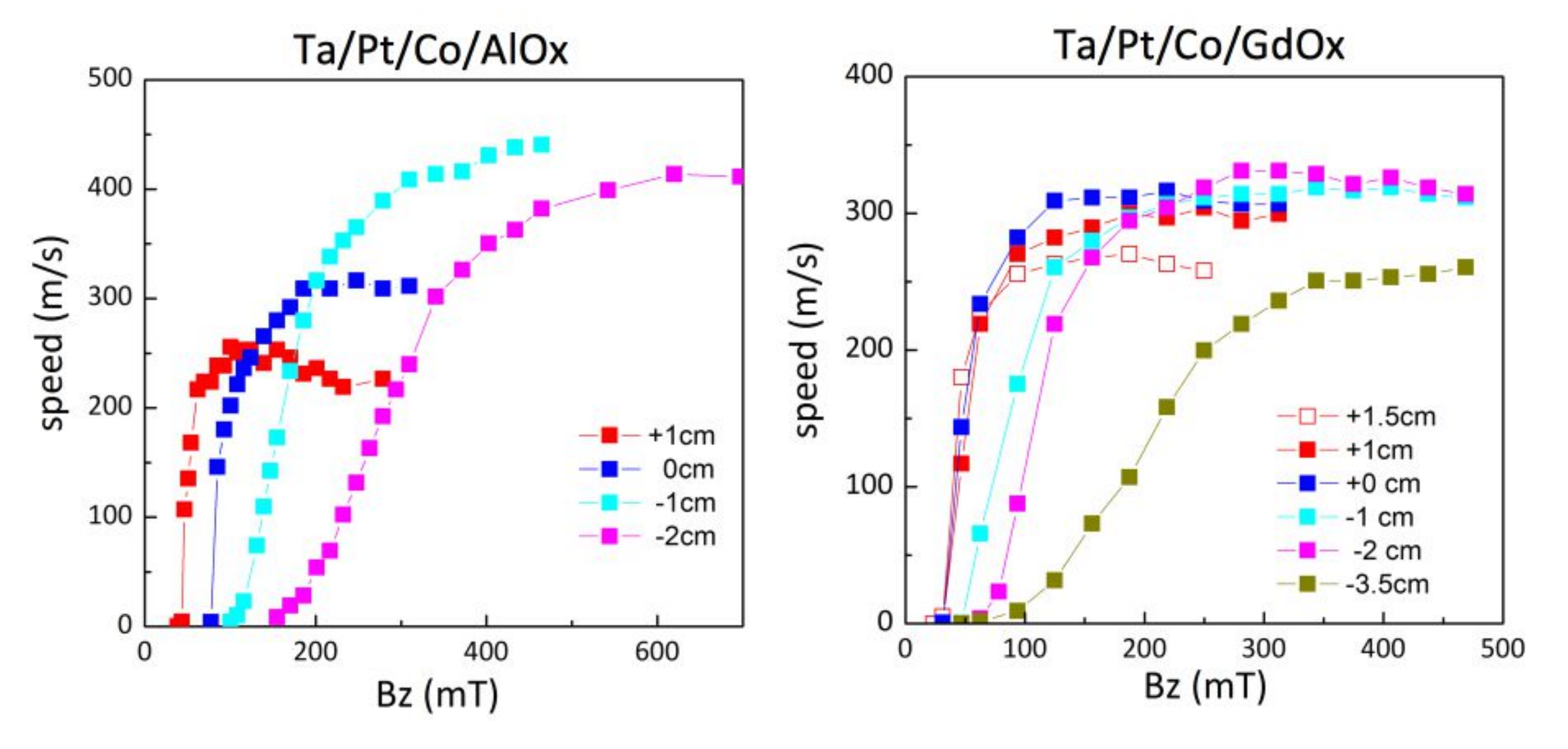}
  \end{center}
\caption{DW speed as a function of out-of-plane field $B_{z}$ for Pt/Co/AlOx and Pt/Co/GdOx for different wedge positions reflecting different oxygen coverage of the Co/MOx interface.}
\label{fig:Figure-2}
\end{figure}

The strength and the sign of the DMI was addressed using two methods based on DW dynamics. The first method,  applied here for the first time for a quantitative estimation of the DMI, relies on the measurement of the DW speed as a function of easy-axis field magnitude $H_{z}$. It was recently shown \cite{Yoshimura2015,Yamada2015,Pham2016}  that in multilayers with strong DMI the field driven DW speeds largely exceed those obtained in symmetric systems for the same fields \cite{Metaxas2007,Pham2016} and that the speed  does not decrease after the Walker field (cancellation of the so-called Walker breakdown), as predicted by 1D micromagnetic simulations. Instead, the speed reaches a plateau that can extend up to very large fields. We have  demonstrated \cite{Pham2016} that the saturation DW speed  can be easily related to the DMI strength \textit{via} the expression:

\begin{equation} \label{eq:vmax}
  v_{sat} = \frac{\pi}{2} \gamma \frac{D}{M_{s}}
\end{equation}

where $D$ is the DMI strength, $M_{s}$ the spontaneous magnetization, and $\gamma $ is the gyromagnetic ratio. This implies that the measurement of the saturation speed, together with that of $M_{s}$, provides a simple way to obtain the strength of the effective DMI interaction in a trilayer system. Since the DMI in these layers has an interface origin, we find  more appropriate to derive the interfacial DMI strength $D_{s}$ from our experiments, so that:

\begin{equation}\label{eq:Ds}
    D_{s}=Dt=2v_{sat} \frac{M_{s} t} {\pi \gamma}
\end{equation}

where $t$ is the thickness of the ferromagnetic layer. This choice has also the advantage to provide a more precise determination of the DMI strength, as it is related to $M_{s}t$, that is measured directly, rather than $M_{s}$ that requires the precise calibration of the ultrathin ferromagnetic layer thickness $t$. The velocity curves, measured for several thicknesses of the AlOx and GdOx layers, are shown in Fig.~\ref{fig:Figure-2}(a-b).

Since the saturation speed is independent on the DMI sign, the sign of the DMI  obtained using the method  proposed by Je \textit{et al}  \cite{Je2013} and used in our previous studies \cite{Vanatka2015,Pham2016}. This consists in measuring the domain wall expansion by an out-of-plane field, in the presence of a constant in-plane magnetic field $H_{x}$ parallel to the DW normal. In systems with DMI and chiral N\'{e}el walls, the DW propagation is anisotropic in the direction of $H_{x}$ and the DW speed is larger/smaller for the domain walls having magnetization parallel/antiparallel to the in-plane field. Using ths approach, we found that the DMI stabilizes anticlockwise spin rotation within the DWs (negative $D$ values) in all our samples.

For some of the samples, we have compared the DMI values extracted from Eq.~\ref{eq:Ds} to those obtained from the measurement of the $H_{z}$-driven  DW speed as a function of  $H_{x}$.  As shown in Fig.~\ref{fig:Figure-3}, the DW speed reaches a minimum  when the applied in-plane field $H_{x}$ compensates the $H_{DMI}$ field that stabilizes the N\'{e}el walls. We can then deduce the average DMI energy density $D$, since
  $H_{DMI} = D /\mu_{0}M_{s} \Delta$
  where $\Delta = \sqrt{A/K_{0}}$, $A$ is the exchange stiffness and $K_{0}$ the effective anisotropy energy. Note that this method presents the disadvantage of requiring the value of the exchange stiffness, a parameter that is in general difficult to evaluate in ultrathin films. In this study we have used $A=16$ pJ/m, a value for which in our previous work on similar samples we found the best agreement with the DMI values measured by BLS. The  $H_{z}$ field driving the DW was always chosen to be beyond the depinning field, so that the complications shown to occur in the creep regime were avoided \cite{Vanatka2015}. The same DMI trends as a function of MOx thickness are obtained with the two methods, with DMI values differing by less that 10\%.  This demonstrates the reliability of the DMI values obtained using Eq. ~\ref{eq:Ds}, which relies only on measured magnetic parameters.

\begin{table*}[t]
\caption{Unit surface magnetization $M_{s}t$, effective anisotropy field  $\mu_{0}H_{K}$, saturation domain wall speed $v_{max}$ , DMI field $\mu_{0}H_{DMI}$, interface DMI energy density $D_{s}$ extracted from $v_{max}$ and from $\mu_{0}H_{DMI}$, for some selected positions along the two wedged samples. }
\label{fig:Table1}
\scriptsize
\begin{center}
\begin{tabular}{  p{2.5cm}  p{1.5cm} p{1.5cm}  p{1.5cm}  p{1.5cm} p{1.5cm}  p{2cm}   p{2cm}     }
\hline \hline \\[-2ex]
sample  & $x$  & $M_{s}t$    & $\mu_{0}H_{K}$   & $v^{max}$  & $\mu_{0}H_{DMI}$      & $D_{s}^{vmax}$ & $D_{s}^{HDMI}$  \\

            & (cm)   & ($10^{-3}$A)      &   (Tesla)  &  (m/s)   & (mT)       & (pJ/m)  & (pJ/m) \\  \\[-2ex]\hline
                                                                                        \\[-2ex]
Pt/Co/AlOx  & -2   & 0.5$\pm$0.1   &  1.8$\pm$0.1  &       410$\pm$20    &        & 0.8$\pm$0.1    &         \\
            & -1.5 & 0.7                &  1.75         &                     &        & 1.06   &         \\
            & -1   & 0.9                &  1.52         &       440           &        & 1.43   &         \\
            & 0   & 1.3                &  0.90         &       310           &        & 1.42   &         \\
            & +1    & 1.4                   &  0.53         &      240           &        & 1.22   &         \\
            &  +2  & 1.5                   &  0.23         &                     &        &        &         \\ \\
            Pt/Co/GdOx  & -3.5 & 0.68   &  1.62   &        255  &  $>$260      & 0.62   &  0.6 $\pm$0.2   \\ 
            & -2     & 0.93  &  1.24   &        330   &  260   & 1.11   &  1.04    \\
            & -1 & 0.97 &  1.05   &        320   &  220   & 1.08   &  1.02    \\
            & 0    & 1.12   &  0.62   &        315   &  190   & 1.34   &  1.42    \\
            & +1   & 1.12   &  0.74   &        300   &  150   & 1.29   &  1.02   \\
            & +1.5 & 1.11   &  0.77   &       270    &  170   & 1.12   &  0.98    \\
                                                                                        \\[-2ex]
\hline
\hline
\end{tabular}
\end{center}
\end{table*}

The interfacial DMI strengths measured for the various wedge positions are shown in Table 1. In Fig.~\ref{fig:Figure-4}, the variation of the interfacial DMI along the AlOx and GdOx wedges is plotted together with the variation of the unit surface magnetization $M_{s}t$ and the interface anisotropy energy density, calculated from $K_{s}=K_{u}t$ with $K_{u}=0.5 \mu_{0}H_{K} M_{s} + 0.5\mu_{0}M_{s}^{2}$.

For both Pt/Co/MOx samples, $M_{s}t$ is larger on the thick MOx side and decreases on the thinner side as the Co-M-O bonds at the interface are substituted by Co-O-M bonds. The interface magnetic anisotropy $K_{s}$  strongly depends on the wedge position. The maximum of $K_{s}$ is observed for an intermediate position, where $M_{s}t$ is maximum or starts decreasing. Coherently, in both trilayer samples, the region with maximum  $K_{s}$ corresponds with the region with maximum coercive field and square hysteresis loops (see Fig.~\ref{fig:Figure-1}).
These results are in agreement with those found in the literature  for a series of Pt/Co/Al samples in which the degree of oxidation of the top Co interface was tuned to optimize the perpendicular magnetic anisotropy (PMA) \cite{Manchon2008a,Manchon2008b,Monso2002,Rodmacq2009}. The onset of PMA was related to the appearance of a significant density of Co-O bondings at the Co/AlOx interface and the PMA was shown to be maximum when the Co top layer was completely covered by oxygen atoms \cite{Manchon2008a,Manchon2008b}.

The main result of this work is that in both samples the interface DMI strength changes along the wedge, and reaches a maximum for the region where the perpendicular magnetic anisotropy is also the largest (see Fig.~\ref{fig:Figure-4}). Since the Pt/Co interface remains unchanged along the wedge (the largest oxide thickness is only 0.3nm thick for a nominal Co thickness of 1nm), the variation of the interface DMI can be ascribed to the modification of the top Co interface only. In both Pt/Co/AlOx and Pt/Co/GdOx samples, while the oxygen coverage changes continuously along the wedge, the DMI strength changes in a \textit{non monotonous} way similarly to the magnetic anisotropy. The optimum coverage of the top Co interface with oxygen not only determines the maximum anisotropy but also optimizes the interfacial DMI. This invokes similar microscopic mechanisms for the variation of $K_{s}$ and DMI.

Some recent theoretical works invoke the presence of DMI at the interface between a ferromagnetic layer (Co or Fe) and an electronegative atom (oxygen or carbon). The DFT calculations of Belabbes \textit{et al.} \cite{Belabbes2017} consider the case of Ir(001)/Fe(1ML)/O and show that the DMI can be tuned by controlling the oxygen coverage. In this extreme case where the FM is only 1ML thick and the FM layer cannot be seen as having two separate interfaces, it is shown that the DMI strength and sign are controlled by the hybridization of the Co-$3d$ and HM-$5d$ electronic bands \cite{Belabbes2016,Belabbes2017}. The charge transfer between Co and O, observed for example by Daalderop \textit{et al.} \cite{Daalderop1994} can modify the Co-HM distances and modify the $3d$-$5d$ hybridisation, therefore changing the $3d$-DOS at the Fermi level and the DMI.

The \textit{ab initio} calculations in Ref. \cite{Boulle2016} consider a Pt(111)/Co(3ML)/MgO trilayer. The layer resolved calculations show that although the main contribution to the DMI is situated at the Pt/Co interface, a significant DMI is also present at the Co/O interface. The latter DMI has the same sign as that of the Pt/Co interface and therefore enhances the total DMI. This study considering a 3ML thick FM layer has to be distinguished from \cite{Belabbes2017} in the sense that the oxygen coverage does not simply leverage a change of the DMI at the HM/FM interface by modifying the electronic structure, but also promotes a (separate) DMI contribution centered at the Co/O interface.

In the systems studied here, where the Co layers are thicker than 3ML, the Pt/Co and Co/MOx interfaces can then be considered as contributing independently to the DMI. We then attribute the variation of the total DMI as a function of oxygen coverage to a DMI contribution located at the Co/O interface, whose strength varies and reaches a maximum, like the PMA, for the optimum oxygen coverage. This conclusion is in line with the first principle calculations of \cite{Yang2011} that studied the variation of the PMA of a Fe/MgO (and Co/MgO) bilayer for different oxidation degrees of the Fe (Co) interface. Despite the weak spin orbit interaction at FM/O interfaces, oxidation is shown to play an essential role in the PMA. The maximum anisotropy is obtained for the oxygen-terminated interfaces, while the PMA decreases for over- or under-oxidized interfaces. The origin of PMA is attributed to the overlap between O-$p_{z}$ and transition metal $3d$ orbitals (optimized for O atoms on top of the FM) that give stronger spin-orbit coupling-induced splitting around the Fermi level for perpendicular magnetization orientation. The precise mechanism by which the orbital hybridization at the Co/O interface also tunes the DMI is still an open question and this finding should stimulate new theoretical studies.

In conclusion, we have shown experimentally that in Pt/Co/AlOx and Pt/Co/GdOx samples the Co/O interface is the source of a DMI contribution, adding to that of the Pt/Co interface, and that can be optimized by controlling the oxygen concentration at the Co/MOx interface. The microscopic origin of such effect is probably found in the details of the hybridization of the Co-$3d$ and O-$2p$ orbitals, which also allows to optimize the PMA. We can anticipate that in these Pt/Co/MOx trilayers the DMI may be manipulated with the application of an electric field, similarly to what we have shown for the effect of E-field on magnetic anisotropy \cite{BernandMantel2012}. This may provide a smart way to engineer future spin-orbitronic devices for controllable chiral domain walls or skyrmion dynamics.

S.~P. and J.~V. acknowledge the support of the Agence Nationale de la Recherche, project ANR-14-CE26-0012 (ULTRASKY). B. Fernandez, Ph. David and E. Mossang are acknowledged for their technical help. D.S.C was supported by a CNPq Scholarship (Brazil). S.~P. thanks A. Thiaville for fruitful discussions.

\end{document}